\documentclass[aps,twocolumn]{revtex4}
\usepackage{newlfont}
\usepackage{amssymb}
\usepackage{amsfonts}
\usepackage{amsmath}

\usepackage{graphicx}
\usepackage{bm}

\usepackage{graphicx}
\usepackage{epsfig}
\usepackage{newlfont}
\usepackage{amssymb}
\usepackage{amsfonts}
\usepackage{amsmath}
\usepackage{graphicx}
\usepackage{bm}

\usepackage{amsthm}

\begin{document}



\title{Entanglement Mean Field Theory and the Curie-Weiss Law}

\author{Aditi Sen(De) and Ujjwal Sen}

\affiliation{Harish-Chandra Research Institute, Chhatnag Road, Jhunsi, Allahabad 211 019, India}

\begin{abstract}
The mean field theory, in its different hues, form one of the most useful
tools for calculating the single-body physical 
properties of a many-body system.
It provides important information, like critical exponents, of the systems that do not yield 
to an exact analytical treatment. Here we propose an entanglement mean field theory (EMFT) 
to obtain the behavior of the two-body physical properties of such systems. We apply this theory to predict  the 
phases in paradigmatic strongly correlated systems, viz. the transverse anisotropic XY, the  transverse XX, 
and the Heisenberg models. We find the critical exponents of different physical quantities in the EMFT limit, and in the case of the Heisenberg model, we 
obtain the Curie-Weiss law for correlations. While the exemplary models have all been
chosen to be quantum ones, classical many-body models also render themselves to such a treatment, at 
the level of correlations.    
\end{abstract}

\maketitle



\section{Introduction}

Characterizing many-body systems by understanding their different phases is a key issue in physics. 
However, 
it is only in a few cases that exact analytical techniques can be applied
\cite{eita-prothhom}.
It is therefore crucial to have approximate methods to deal with such systems to predict their different 
physical properties \cite{eita-dwitiyo}. 
A very useful method is to use mean field theory (MFT) \cite{eita-dwitiyo, eita-MFT},
which 
renders the many-body physical system into one with a single particle. 
MFT allows one 
to 
predict the single-body physical properties, like magnetization, susceptibility,
 of the system. Its importance lies in the fact that these MFT-reduced one-body properties
can correctly predict the thermal fluctuation- and quantum fluctuation- driven phase
transitions
of the system, as well as the critical exponents of the such single-body physical quantities \cite{komlalebu1, komlalebu2}.

Useful as it is, there are important limitations of a mean field theory \cite{eita-MFT}, a 
first
being that it is 
not possible to predict the multi-party physical properties, like entanglement, of the system by using this theory. 

In this paper, we propose an ``entanglement mean field theory'' (EMFT), which transforms an 
interacting 
many-body physical system into a two-body one, while still retaining certain footprints 
of the interactions in the many-body parent, using which it is possible to calculate the two-body properties, 
like two-point entanglement and two-point correlations, of the system, and predict the critical phenomena in it.
Moreover, it is possible to calculate the critical exponents of two-body properties. 
The theory can be applied for detecting phase transitions driven by thermal fluctuations
as well as quantum fluctuations.
At the same time, both quantum as well as classical models can be treated by EMFT, where in the classical case, this 
will be only until the level of correlations. In this paper, we will only consider the applications
of EMFT to  
quantum systems. We will consider three paradigmatic classes of interacting spin models, viz.
the transverse anisotropic quantum XY (which includes the transverse quantum Ising), 
the 
transverse quantum XX,
and the quantum Heisenberg models. The phase 
transitions of these models, both temperature-induced phase transitions as well as 
zero temperature quantum phase transitions, are faithfully signaled 
by the corresponding EMFT entanglements.  
The Curie-Weiss law of susceptibility is an important prediction of the mean field-reduced Heisenberg model
in the paramagnetic regime \cite{eita-MFT}. 
We indicate the corresponding law for EMFT-reduced correlations by using the 
EMFT-reduced Heisenberg model. 

The entanglement mean field theory 
opens up the possibility of investigating the behavior of entanglement and other
two-body properties of many-body systems, particularly for the ones which does not lend themselves to 
an analytical treatment.  
At the fundamental level, 
this forms, potentially,
  an important link 
 between 
many-body physics and quantum information science  \cite{6}.

\section{The Entanglement Mean Field Theory: XY model}

Before presenting the entanglement mean field theory, let us briefly remind ourselves the mean field theory. 
Consider the transverse quantum anisotropic XY model with nearest-neighbor interactions,
 described by the Hamiltonian
\begin{eqnarray}
\label{hamil-xy}
H^{XY}= -\frac{J}{2}\sum_{\langle {\vec{i}}{\vec{j}} \rangle}\left[(1+\gamma)\sigma_x^{\vec{i}} \sigma_x^{\vec{j}} 
                                                                  + (1-\gamma)\sigma_y^{\vec{i}} \sigma_y^{\vec{j}}\right] 
                                                                      - h \sum_{\vec{i}} \sigma_z^{\vec{i}}, \nonumber \\
\end{eqnarray}
which represents a system of interacting spin-1/2 particles on a \(d\)-dimensional cubic lattice.
The coupling strength \(J/2\) is positive, the anisotropy \(\gamma \in (0,1]\), and the 
transverse field strength \(h\) is also 
positive. 
\(\sigma_x\), \(\sigma_y\), and \(\sigma_z\) are the Pauli matrices for 
the spin degree of freedom of a 
spin-1/2 particle.  \(\langle {\vec{i}} {\vec{j}}\rangle\) indicates that the corresponding sum runs over 
nearest neighbor lattice sites only.
The mean field theory consists in assuming that a particular spin, say \(\vec{i_0}\), is special, and 
replacing all other spin operators by their mean values. 
Denoting the mean values of the spin operators \(\sigma_x\), \(\sigma_y\), \(\sigma_z\) as 
\(m_x\), \(m_y\), \(m_z\), respectively, this leads to an MFT Hamiltonian \( H_{MFT}^{XY} \) \cite{ebar-baRi-jabo, Kol}.
We then solve the self-consistency equations (mean field equations)
\begin{equation}
m_x=\mbox{tr}\left(\sigma_x \rho_\beta\right), \quad m_y=\mbox{tr}\left(\sigma_y \rho_\beta\right), 
\end{equation} 
where \(\rho_\beta\) is the mean field canonical equilibrium state 
\(\exp(-\beta H^{XY}_{MFT})/\mbox{tr}(\exp(-\beta H^{XY}_{MFT}))\),
for \(m_x\) and \(m_y\), substitute them in \(H^{XY}_{MFT}\) and \(\rho_\beta\), and we are then ready to 
find the single-body physical properties of the system in the mean field limit.
 Here \(\beta = \frac{1}{k_B T}\), with \(T\)
denoting temperature on the absolute scale, and \(k_B\) denoting the Boltzmann constant.

The entanglement mean field theory begins by replacing an identity, on a site that is neighboring the interacting spins of a two-spin interaction term,
by a square of the Pauli matrix that is involved in the interaction, for all the two-spin interaction 
terms in the Hamiltonian. [An averaging needs to be done for two-spin interactions involving two Pauli matrices.]
Therefore, the term \(\sigma_x^{k-1,l}\sigma_x^{k,l}\) in a Hamiltonian on a two-dimensional square lattice can be replaced by 
\(\sigma_x^{k-1,l}\sigma_x^{k,l}\sigma_x^{k+1,l}\sigma_x^{k+1,l}\). The latter can be re-written as \(AB\), with
\(A=\sigma_x^{k,l}\sigma_x^{k+1,l}\), and \(B=\sigma_x^{k-1,l}\sigma_x^{k+1,l}\).
We then assume 
%
%
%
that a certain pair of two neighboring spins are ``special''. 
We then
replace 
the non-special two-spin interactions (with nearby spins) in all the interaction terms by their mean values. 
In a \(d\)-dimensional cubic lattice, there will be \(2\nu\) terms (with \(\nu +1 = 2d\) being the 
coordination number of the lattice), that will have the special pair, along with 
two other spins in two other lattice sites. The EMFT-reduced Hamiltonian, for the transverse quantum ferromagnetic
XY model on a \(d\)-dimensional cubic lattice, will therefore be
\begin{eqnarray}
\label{emft-asol}
{\cal H}^{XY}_{EMFT} = -J\nu \left[ (1+\gamma)C_{xx}\sigma_x^{\vec{i}} \sigma_x^{\vec{j}}  
                      + (1-\gamma)C_{yy}\sigma_y^{\vec{i}} \sigma_y^{\vec{j}} \right] \nonumber \\ 
                          - h\left[ \sigma_z^{\vec{i}} + \sigma_z^{\vec{j}} \right)],
\end{eqnarray}
where we have ignored the terms in the Hamiltonian which will not contribute to the EMFT equations below, and where 
we have assumed that the neighboring lattice sites \({\vec{i}}\) and \({\vec{j}}\) are special.
\(C_{xx}\) and \(C_{yy}\) are respectively the \(xx\) and \(yy\) correlators, and 
the self-consistency equations (EMFT equations) are
\begin{equation}
C_{xx}= \mbox{tr} \left( \sigma_x^{\vec{i}} \sigma_x^{\vec{j}} \varrho_\beta \right), \quad 
C_{yy}= \mbox{tr} \left( \sigma_y^{\vec{i}} \sigma_y^{\vec{j}} \varrho_\beta \right),
\end{equation}
where \(\varrho_\beta\) is the canonical equilibrium state 
\(\exp(-\beta {\cal H}^{XY}_{EMFT})/\mbox{tr}(\exp(-\beta {\cal H}^{XY}_{EMFT}))\). The EMFT equations  are to be solved 
for \(C_{xx}\) and \(C_{yy}\), and substituted in \({\cal H}^{XY}_{EMFT}\) and 
\(\varrho_\beta\). We can then calculate the two-particle physical properties of the physical system described by 
\(H^{XY}\) in the EMFT limit, and look for possible phase transitions of the system.


A similar formalism works,  with  slight modifications, for classical spins, higher quantum spins, more complex lattices, etc.
Also, both the mean field theory as well as the EMFT has been described for the ferromagnetic cases. The antiferromagnetic 
case requires some modifications in the mean field theory, and correspondingly some changes in the EMFT. These will not be 
discussed in this paper.

\subsection{The Transverse Ising model}

For simplicity, let us consider the transverse quantum Ising model (\(\gamma =1\)). 
This is the simplest model which exhibits a quantum phase transition (at zero
temperature) \cite{komlalebu2, Kol}, that has been experimentally observed \cite{Ising_experiment}.  
for which the EMFT equation
reads
\begin{eqnarray}
C_{xx}=  \frac{2}{{\cal Z}^{Ising}_{EMFT}}\Big[\sinh(2\beta \nu C_{xx} J ) 
                   + \frac{\nu C_{xx} J}{\Gamma} \sinh(2\beta \Gamma)\Big],
\end{eqnarray}
where the corresponding EMFT Ising partition function is given by 
\({\cal Z}^{Ising}_{EMFT} = 2\cosh(2\beta \nu C_{xx} J )
                   +  2\cosh(2\beta \Gamma)\), with \(\Gamma = \sqrt{\nu^2 C_{xx}^2 J^2 + h^2}\).

We are now ready to present the EMFT phase diagram, where we use 
the \(xx\) correlator, \(C_{xx}\), as the order parameter of the many-body system. 
Let us first look at the \(T=0\) picture, where transitions are driven by quantum fluctuations only. 
The zero temperature EMFT \(C_{xx}\) is given by 
\begin{equation}
\label{snajh-ghurte-jabe-bolchhe}
C_{xx} = \sqrt{1-\frac{h^2}{\nu^2 J^2}},
\end{equation}
for \(h\leq h_c \equiv \nu J\). It is vanishing otherwise. 
%
Let us now fix our attention on the other extreme: the behavior of the system with respect to temperature for zero field, in which case the 
EMFT equation reduces to 
\begin{equation}
C_{xx}= \tanh \left( 2\beta \nu C_{xx} J \right),
\end{equation}
whereby a temperature-driven phase transition is obtained at 
\(T=T_c \equiv 2\nu J/k_B\). 
The complete phase diagram can be seen in Fig. 1, which contains these extreme cases as special instances. 

\begin{figure}[h!]
\label{fig-chhobi-ek}
\begin{center}
\epsfig{figure= 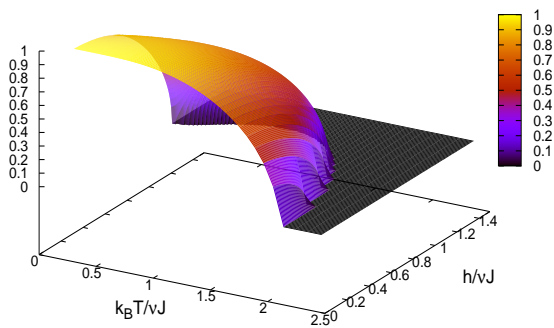, height=.35\textheight,width=0.35\textwidth, angle = -90}
\caption{
The \(xx\) correlator, \(C_{xx}\), of the transverse Ising model in the EMFT limit. 
The temperature-driven phase transition at \(T=T_c\) and the quantum phase transition at \(h=h_c\) are clearly visible on the respective axes. 
All the three axes represent dimensionless quantities.
}
\end{center}
\end{figure}

It is possible to calculate the critical exponents in the EMFT limit. The critical exponent for the
EMFT \(xx\) correlator can be calculated as follows, which we find for both 
the temperature-driven phase transition 
on the \(h=0\) axis, and for the quantum phase transition on the \(T=0\) axis.
In the zero temperature scenario, 
the critical exponent is \(\frac{1}{2}\), as can be found by using Eq. (\ref{snajh-ghurte-jabe-bolchhe}). In the 
zero field case, we can perform an expansion of the equation, 
\begin{equation}
 2\beta \nu C_{xx} J = \tanh^{-1} C_{xx},
\end{equation}
around \(T=T_c\), and the critical exponent is again \(\frac{1}{2}\).
In a similar fashion, one can obtain the critical exponent for the EMFT energy gap to be \(\frac{1}{2}\). Note that the same exponent is obtained 
for the gap in the MFT limit \cite{monta-uttapam-uttapam-korchhe}.

The entanglement mean field theory can not only predict the correlations of the system, but one can also 
study the behavior of its entanglement. Apart from its fundamental importance, entanglement is known to be the
basic ingredient in quantum information tasks \cite{LIC}.  
We will quantify the 
entanglement of a two-party quantum state \(\varrho\), by its logarithmic negativity \cite{Jeevan-komal}, defined as 
\begin{equation}
E(\varrho) = \log_2 \parallel \varrho^{T_1}\parallel_1, 
\end{equation}
 where \(\parallel \cdot \parallel\) denotes the trace norm of its argument, and 
\(\varrho^{T_1}\) denotes the partial transpose of \(\varrho\) with respect to one of the two parties forming the state \(\varrho\). 
The behavior of entanglement in the EMFT canonical state with respect to temperature and applied field, in the transverse Ising model, 
is seen in Fig. 2.
\begin{figure}[h!]
\label{fig-chhobi-dui}
\begin{center}
\epsfig{figure= 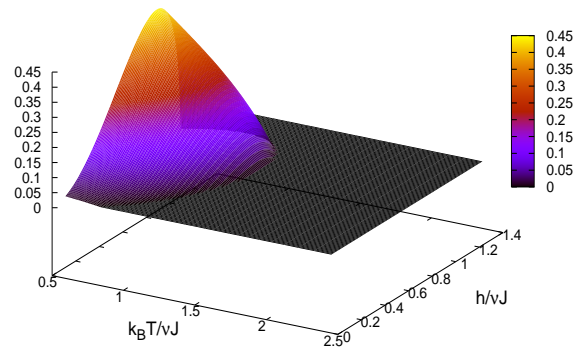, height=.35\textheight,width=0.35\textwidth, angle = -90}
\caption{
EMFT entanglement in the transverse Ising model, plotted against temperature and field strength. 
The base axes represent dimensionless system parameters, while the vertical axis represent entanglement, measured in 
ebits. 
}
\end{center}
\end{figure}

%
%


The other members of the class of Hamiltonians given in Eq. (\ref{hamil-xy}) have a similar behavior with respect to their two-body physical properties in the 
EMFT limit.  We can compare this with the fact that they fall in the same universality class \cite{komlalebu2}. 
A different universality class is considered in the succeeding section.

\section{The EMFT-reduced XX model}

The transverse field quantum XX model  on a \(d\)-dimensional cubic lattice is described by the Hamiltonian
\begin{eqnarray}
\label{hamil-xx}
H^{XX}= -\frac{J}{2}\sum_{\langle {\vec{i}} {\vec{j}} \rangle}\left[\sigma_x^{\vec{i}} \sigma_x^{\vec{j}} 
                                                                  + \sigma_y^{\vec{i}} \sigma_y^{\vec{j}}\right] 
                                                                      - h \sum_{\vec{i}} \sigma_z^{\vec{i}}, 
\end{eqnarray}
where \(J\) and \(h\) are positive.
The physical importance of the Hamiltonian includes that it can be 
obtained from the Bose Hubbard Hamiltonian for hard-core boson limit, by suitably associating the 
bosonic creation and annihilation operators with the Pauli matrices \cite{komlalebu2}. 
The corresponding EMFT-reduced Hamiltonian is 
\begin{eqnarray}
\label{emft-asol-xx}
{\cal H}^{XX}_{EMFT} = -J\nu C \left[\sigma_x^{\vec{i}} \sigma_x^{\vec{j}}  
                      + \sigma_y^{\vec{i}} \sigma_y^{\vec{j}} \right] 
                          - h\left[ \sigma_z^{\vec{i}} + \sigma_z^{\vec{j}} \right)],
\end{eqnarray}
where \(C\) is the \(xx\) (which is same as the \(yy\)) correlator of the system, and where we have supposed that the neighboring lattice sites \({\vec{i}}\) and 
\({\vec{j}}\) are special.

The EMFT equation in this case reads
\begin{equation}
C=2 \sinh \left( 2 \beta \nu C J  \right)/ {\cal Z}^{XX}_{EMFT},    
\end{equation}
where the EMFT XX partition function is given by 
\({\cal Z}^{XX}_{EMFT} = 2 \cosh \left( 2 \beta \nu C J  \right) + 2  \cosh \left( 2 \beta h  \right) \).
Solving for \(C\) from the EMFT equation, we can subsequently find other physical quantities of the system. In particular, the entanglement of the system, 
as quantified by its logarithmic negativity, is given in Fig. 3.
\begin{figure}[h!]
\label{fig-chhobi-tin}
\begin{center}
\epsfig{figure= 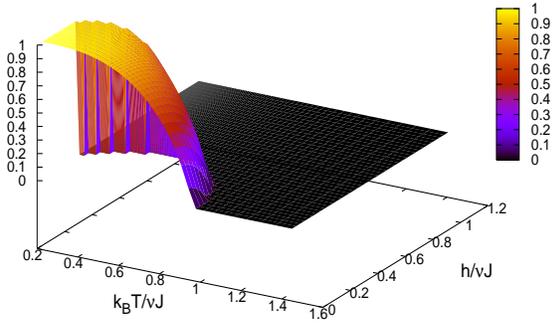, height=.35\textheight,width=0.35\textwidth,  angle=-90}
\caption{
EMFT entanglement in the transverse XX model. The plot shows the behavior of entanglement of the canonical state of the EMFT-reduced transverse XX model 
with respect to temperature and applied field.   The base axes represent dimensionless system parameters, while the vertical axis 
represent entanglement, measured in ebits. 
}
\end{center}
\end{figure}

It is also possible to write an EMFT equation for the ground state of the EMFT-reduced XX Hamiltonian, solving which we find 
\begin{eqnarray}
 C=1, \mbox{ for } \frac{h}{J \nu} <1 \nonumber \\
 C=0, \mbox{ for } \frac{h}{J \nu} >1. 
\end{eqnarray}
It is interesting to compare this result with the fact that the physical system represented by the Hamiltonian 
\(H^{XX}\), in the one-dimensional case, undergoes a Mott insulator to superfluid 
transition at \(h/J=1\) \cite{Aditi-r-compu-kharap, komlalebu2}.

\section{EMFT and the Heisenberg model: A Correlation Curie-Weiss Law}

Investigations in magnetism in solids very often starts off by using the Heisenberg Hamiltonian \cite{Snajh-ekhono-school-e}, which, for 
a \(d\)-dimensional cubic lattice, is given by 
\begin{eqnarray}
\label{hamil-heis}
H^{HD}= -J\sum_{\langle {\vec{i}}{\vec{j}} \rangle}\left[\sigma_x^{\vec{i}} \sigma_x^{\vec{j}} 
                                                                  + \sigma_y^{\vec{i}} \sigma_y^{\vec{j}} + \sigma_z^{\vec{i}} \sigma_z^{\vec{j}}\right] 
\nonumber \\
                               - h \mu \sum_{\vec{i}} \left[ \sigma_x^{\vec{i}} + \sigma_y^{\vec{i}} + \sigma_z^{\vec{i}} \right], 
\end{eqnarray}
where \(J\), \(h\), and \(\mu\) are positive. 
An important conclusion of the MFT treatment of this model is the Curie-Weiss Law, which predicts the behavior of magnetization of the physical system 
in its paramagnetic phase. We will see that it is possible to extract a similar law for the correlations in the system by solving the 
corresponding EMFT Hamiltonian. 

The EMFT-reduced Heisenberg Hamiltonian is given by 
\begin{eqnarray}
\label{emft-asol-heis}
{\cal H}^{HD}_{EMFT} = -2J\nu C \left[ \sigma_x^{\vec{i}} \sigma_x^{\vec{j}}  
                      + \sigma_y^{\vec{i}} \sigma_y^{\vec{j}} + \sigma_z^{\vec{i}} \sigma_z^{\vec{j}} \right] \nonumber \\ 
                          - h \mu\left[ \sigma_x^{\vec{i}} + \sigma_x^{\vec{j}} + \sigma_y^{\vec{i}} + \sigma_y^{\vec{j}} +\sigma_z^{\vec{i}} + \sigma_z^{\vec{j}} \right)],
\end{eqnarray}
where \(C\) is the \(zz\) (which is the same as the \(xx\) and \(yy\)) correlation of the system, and where
we have supposed that the lattice sites 
\(\vec{i}\) and \(\vec{j}\) are special, for constructing the EMFT Hamiltonian. 
The EMFT equation in this case is 
\begin{equation}
\label{khide-peyechhe}
 C = \frac{2}{3 {\cal Z}^{HD}_{EMFT}} \left[  \sinh \Gamma_C - \exp(- \Gamma_C) + \exp \Gamma_C \cosh \Gamma_h  \right]
\end{equation}
where the EMFT Heisenberg partition function is given by 
\({\cal Z}^{HD}_{EMFT} = 2 \cosh \Gamma_C + 2\exp \Gamma_C \cosh \Gamma_h\), with 
\(\Gamma_C=2 \beta \nu C J\) and \(\Gamma_h = 2 \sqrt{3} \beta h \mu\).

The partial derivative of magnetization with respect to the applied field is defined as the susceptibility of the system, and 
the usual Curie-Weiss Law is given for that quantity, for vanishing applied field. We define the correlation susceptibility of the system as the 
partial derivative of the correlation \(C\) with respect to the field strength \(h\). This quantity, as solved from 
the EMFT equation (Eq. (\ref{khide-peyechhe})), for \(h=0\),
is given in Fig. 4. 
Note that the correlation susceptibility clearly signals the onset of the paramagnetic phase of the system at 
\(T=T_c^{FP} \equiv \nu J/k_B\). 
The data obtained for \(\frac{\nu J}{\mu}\frac{\partial C}{\partial h}\big|_{h=0}\) can be fitted to the curve 
\(\left(\frac{\nu J}{k_B}\right)^\lambda \alpha/(T-T_c^{FP})^\lambda\), for \(T>1.2 T_c^{FP}\), and it gives the optimal values of the curve 
parameters as \(\alpha= 0.06895\) and \(\lambda=1.08162\). The corresponding mean square error is \(1.2 \times 10^{-4}\).

\begin{figure}[h!]
\label{fig-chhobi-char}
\begin{center}
\epsfig{figure= 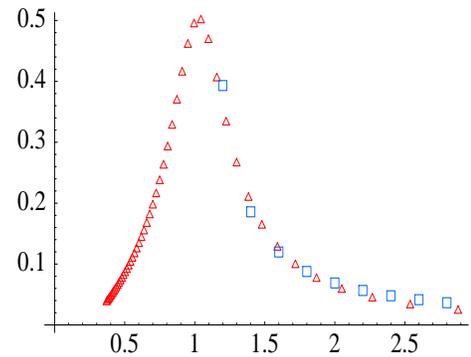, height=.2\textheight,width=0.35\textwidth}
\caption{
Curie-Weiss Law of EMFT-reduced correlation in the Heisenberg model. The partial derivative of the \(zz\) correlation (which is same as the \(xx\) and \(yy\) ones)
with respect to the field \(h\) is plotted, as red triangles, against temperature, for \(h=0\).  More precisely, 
\(\frac{\nu J}{\mu}\frac{\partial C}{\partial h}\big|_{h=0}\) is plotted on the vertical axis as a function of \(\frac{k_B T}{\nu J}\)
on the horizontal axis.  Both axes
represent dimensionless quantities. The blue squares represent a fit of the plotted data with the function 
\(\left(\frac{\nu J}{k_B}\right)^\lambda \alpha/(T-T_c^{FP})^\lambda\), for \(T>1.2 T_c^{FP}\). See text for further details.
}
\end{center}
\end{figure}


\section{Conclusions}
The mean field theory is a useful tool to obtain important information about single-body physical 
quantities of many-body systems, especially the ones which are not tractable analytically. 
We have presented an entanglement mean field theory that can be used to obtain information about 
two-body physical quantities of many-body systems. 
The theory predicts the phase diagram of the physical system, and the critical exponents of their two-body quantities, as we have shown for several 
important classes of many-body systems. In particular, we have derived a Curie-Weiss Law for correlations in the Heisenberg spin model.

\acknowledgments
We acknowledge partial support from the Spanish MEC (TOQATA (FIS2008-00784)).

\end{document}